\documentclass[10pt]{blois07}
\usepackage{epsf,epsfig}
\usepackage{graphicx}
\usepackage{./cite,./mcite}

\begin{document}
\title{Blois07/EDS07 \\ Exclusive vector meson electroproduction }
\author{D.Yu. Ivanov}
\institute{Sobolev Institute of Mathematics, 630090 Novosibirsk,
Russia} \maketitle

\begin{abstract}
We discuss exclusive vector meson electroproduction within the
QCD collinear factorization framework. In Bjorken kinematics
the amplitude factorizes in a convolution of the nonperturbative
meson distribution amplitude and the generalized parton densities
with the perturbatively calculable hard-scattering amplitudes,
which are presently known to next-to-leading order (NLO).
At small $x_{\rm B}$ NLO corrections are very large. It is related to
appearance of BFKL type logarithms in the hard-scattering amplitudes,
that calls for a resummation of these effects at higher orders.
Here we report the first results of such resummation.

\end{abstract}

\section{Introcuction}

The process of
elastic vector meson electroproduction on a nucleon,
\begin{equation}
\gamma^*(q)\, N(p) \to V(q^\prime)\, N(p^\prime)
\, ,
\label{process}
\end{equation}
where $V=\rho^0,\,  \omega, \, \phi $, was studied in many fix target
and in HERA collider experiments.
On the theoretical side, the large negative virtuality of the photon,
$q^2=-Q^2$, provides a hard
scale for the process which justifies the application of QCD factorization
methods that allow to separate the contributions to the amplitude
coming from different scales.
The factorization theorem \cite{Collins:1996fb} states that in a scaling limit,
$Q^2\to \infty$ and $x_{B}=Q^2/2(p\cdot q)$ fixed, a vector meson is
produced
in the longitudinally polarized state by the longitudinally polarized
photon
and that the amplitude of the
process (\ref{process}) is given by a convolution of
the perturbatively calculable hard-scattering amplitudes $C^i$, the
nonperturbative
meson distribution amplitude (DA) $\phi_V$, and the generalized parton densities (GPDs) $H^i$.
\begin{equation}
\label{factor}
A= \sum\limits_{i=q,g}\int\int\ dx \ dz\ {H^i(x,\xi,t,\mu_F)} \ C^i(x,z,\mu_F) \
{\phi_V(z,\mu_F)} \, ,
\end{equation}
where
$
\xi={x_B}/({2-x_B})$ is the skewness variable, $t=(p-p^\prime)^2$ and $\mu_F$
is a factorization scale.
GPDs encode important information on hadron structure, including aspects
that cannot be deduced directly from experiment, like the transverse spatial distribution of partons
and their orbital angular momentum, for more details see
\cite{Mueller:1998fv,*Ji:1996ek,*Radyushkin:1996nd,*Burkardt:2000za}.

Deeply virtual Compton process (DVCS) provides the theoretically cleanest access to GPDs.
Recently two-loop effects were incorporated into the analysis of DVCS \cite{Kumericki:2007sa}.
A theoretical description of exclusive meson production is more involved since it includes
an additional nonperturbative quantity, a meson DA.
The primary motivation for the strong interest in this process
(and in the similar process of heavy quarkonium production) is that
it can serve to constrain the gluon density in a nucleon.
Indeed, in vector meson production case the gluon GPD enters the description
already at the leading order (LO) in the strong coupling $\alpha_s$, whereas in DVCS it appears first at NLO,
and, like in inclusive DIS, is accessible only through scaling violation.

\section{NLO corrections}

The hard-scattering amplitudes for process (\ref{process}) were calculated at NLO in \cite{Ivanov:2004zv},
and for exclusive heavy quarkonium photoproduction in \cite{Ivanov:2004vd}.
The analysis of NLO effects showed that in kinematics typical
for the HERA collider experiments, $x_B\sim 10^{-3}$, the NLO corrections are huge even for
really large values of hard scales $\sim 30 \, \rm{GeV}^2$.
If the factorization scale is chosen close to the value of a hard scale, $\mu_F\sim Q$,
the corrections have opposite signs in comparison to the Born term.
Which may lead to the change of signs of
the imaginary and the real parts of the amplitude within phenomenologically relevant interval of $x_B$.
Besides, the factorization and renormalization scale uncertainties were found being very large.

Recently these findings were confirmed in \cite{Diehl:2007hd}, where very detailed analysis of the cross
sections and the transverse target polarization asymmetries in exclusive
meson production was performed both for small and larger values of $x_B$,
typical for fixed-target experiments. For the fixed target kinematics it seems that NLO corrections
start to be under control, though their values are still large at presently available values of $Q^2$.
For the transverse target polarization asymmetries the situation is better, in some cases.

Going back to small $x_B$, why NLO corrections are large in this case?
The inspection of NLO hard-scattering amplitudes shows that
the imaginary part of the amplitude
dominates and that the  leading contribution to the NLO correction originates from the
broad integration region $\xi\ll x\ll 1$, where the gluonic part approximates ($N_c=3$ is a number of colors)
\begin{equation}
 Im A^g \sim
\int\limits^1_{0}\frac{ dz \, \phi_V(z)}{z(1-z)}\Biggl[
H^g(\xi,\xi,t)
+\frac{\alpha_s N_c}{\pi}
\ln\left(\frac{Q^2z(1-z)}{\mu_F^2}\right)
\int\limits^1_{2\xi} \frac{dx}{x}  H^g(x,\xi,t)
\Biggr] \, .
 \label{appr}
\end{equation}
Given the behavior of the gluon GPD at small $x$, $H^g(x,\xi)\sim x g(x)\sim const$,
we see that NLO correction is parametrically
large, $\sim \ln(1/\xi)$, and negative unless one chooses the
value of the factorization scale sufficiently lower than the kinematic
scale. For the asymptotic form of meson DA, $\phi^{as}_V(z)=6z(1-z)$,
the last term in (\ref{appr}) changes the sign at $\mu_F=\frac{Q}{e}$,
for the DA with a more broad shape this happens at even lower values of
$\mu_F$. Similar, $\ln(1/\xi)$ enhanced, contribution appears also in the quark singlet channel.

The partonic momentum fraction $x$ is related to the Mandelstam energy variable $\hat s$ of the partonic
subprocess $x/\xi\sim \hat s/Q^2$. The leading part of NLO partonic amplitude (proper normalized)
grows as the first power of energy, $ x\sim  \hat s$, whereas at LO partonic amplitude
behaves like a constant at large $\hat s$. The reason for this difference is the appearance, starting from NLO,
of partonic diagrams with the gluon exchange in the $t-$ channel, see Fig. 1.
At LO one has only diagrams with the quark exchange, both for the gluon and quark channels.

\begin{figure}[h!]
\begin{center}
\epsfig{figure=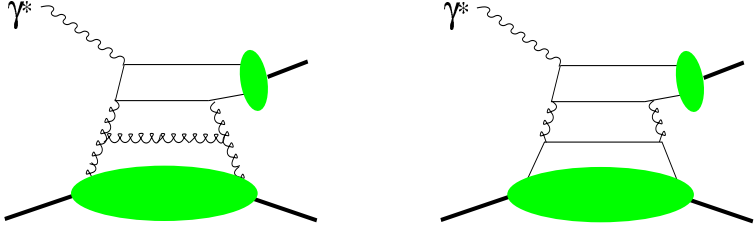,width=10.cm,height=3cm}
\end{center}
\caption{NLO diagrams with $t-$ channel gluon exchange; the gluon and the quark GPD contributions.}
\end{figure}

At higher orders the diagrams with gluon $t-$ channel exchange give contributions to the
amplitudes of partonic subprocesses enhanced,  for $n$ loops, as $\alpha_s^n\log^{n-1} x/x $.
In its turn, these terms inserted in the factorization formula will produce
large contributions $\sim \alpha_s^n\log^{n}(1/ \xi)$ to the process amplitude,
where each power of the strong coupling is compensated
by the same power of a large logarithm of energy. It is a natural idea to resum
these enhanced at small $x_B$ contributions using the
BFKL approach \cite{Kuraev:1977fs,*Balitsky:1978ic}.

\section{High energy resummation}

The central point in this high energy resummation is to perform it consistently,
without spoiling the all-order factorization of collinear singularities.
A care should be taken of the factorization scheme used at the factorization of the process amplitude
(\ref{factor}) in terms of GPDs and
hard-scattering amplitudes.
The higher order terms of the hard-scattering amplitudes derived within the high energy approximation
(BFKL approach) can be supplemented by the knowledge of
hard-scattering amplitudes calculated exactly at fixed order.
Then, one can use them together in factorization formula (\ref{factor}) without double counting.

For inclusive hard processes, heavy quark production and DIS, the method of such high energy resummation
was elaborated in \cite{Catani:1990xk,*Catani:1994sq}.
It is based on Curci, Furmanski and Petronzio approach \cite{Curci:1980uw}
to separation of collinear singularities. The amplitudes on a parton (quark, gluon) target are considered
in $D=4+2\epsilon$ non-integer dimentions. That separates automatically the leading twist.
Collinear singularities appear in this approach as $1/\epsilon^n$ poles,
these poles are absorbed into a definition of parton densities. Another essential ingredients of
\cite{Catani:1990xk,*Catani:1994sq} method is an analysis of Mellin moments,
high-energy terms in Mellin moment space $N$ look like singularities
$\left(\alpha_s/N\right)^n$
at $N\to 0$,
and a consideration of BFKL equation in $D=4+2\epsilon$ dimensions.

We found\footnote{D.Yu.Ivanov, R. Kirschner and A.Papa, in preparation}
that this technique may be directly generalized on the analysis of exclusive non-forward reactions.
Below we present the first results of this study.

Like in DIS the imaginary part of the a
amplitude is given by the sum of quark singlet and gluon contributions
\begin{equation}
\label{ampl}
{\cal I}m A(\xi,t)=\frac{1}{\xi}\int\limits^1_{\xi}dx \left[
D^{(+)}\left(\frac{\xi}{x}\right){H}^{(+)}(x,\xi,t)+\frac{1}{\xi}
D^g\left(\frac{\xi}{x}\right)
{H}^g(x,\xi,t)\right] \, .
\end{equation}
$D^{(+)}$ and $D^g$ are the imaginary parts of the quark and gluon hard-scattering amplitudes.
In difference to forward DIS case the parton densities in (\ref{ampl}) depend on both longitudinal
momentum fractions. Due to that the
Mellin moments of the amplitude do not factorize into the product of the moments
\begin{eqnarray}
\label{moments}
&
D_N(t)=\int\limits^1_0 d\xi \, \xi^N {\cal I}m A(\xi,t)= & \\
&
\int\limits^1_{0}\int\limits^1_{0} du \, dx \, u^{N-1} x^N\,  \left[
D^{(+)}\left(u\right){\cal H}^{(+)}(x,u\, x,t)+\frac{1}{u\, x}
D^g\left(u\right)
{\cal H}^g(x,u\, x,t)\right] \, . &
\nonumber
\end{eqnarray}
Using polynomiality property of GPDs, in particular for the gluon case
\begin{equation}
\int_{0}^1 dx\, x^{n}{ H}^g(x,\eta,t) =
  \sum_{j=0, \rm{even}}^{n} (2\eta)^j A^g_{n+2,j}(t)
  + (2\eta)^{n+2}\, C^g_{n+2}(t) \, ,
\end{equation}
one can show that for the integer odd $N$
\begin{equation}
\label{DN}
D_N(t)=
\sum_{k=0}^{\infty}2^k
\left[
D^{(+)}_{N+k-1}A^q_{N+k+1,k}(t)+
D^{g}_{N+k-2}A^g_{N+k+1,k}(t) \right] \, .
\end{equation}
Which is a sum of moment products (not just a product, as in DIS case).

One can analytically continue (\ref{DN}) from the integer odd $N$ into entire complex $N$ plane.
The high energy asymptotic of the amplitude is related with the behavior of
$D_N(t)$ near unphysical  point $N\to 0$.
One can split (\ref{DN}) into a sum of the singular and the regular at $N\to 0$  parts
\begin{equation}
D_N(t)=C^{(+)}_N q^{(+)}_N(t)+C^{g}_N g_N(t)+D^{\rm{reg}}_N(t)
\end{equation}
The singularities of the sum (\ref{DN}) at $N\to 0$ are due to $k=0$ term only.
Therefore
\begin{equation}
C^{(+)}_N=D^{(+)}_{N-1}, \quad q^{(+)}_N(t)=A^q_{N+1,0}(t)
, \quad
C^{g}_N=D^{g}_{N-2} ,\quad g_N(t)=A^g_{N+1,0}(t) \, .
\end{equation}
Note that
at $t\to 0$, $q^{(+)}_N(t)$ $g_N(t)$ reduce to the moments of usual
parton densities
\begin{equation}
q^{(+)}_N(t)\to  q_N^{(+)}=\int\limits^1_0 dx \, x^N
q^{(+)}(x)\, , \quad
g_N(t)\to  g_N=\int\limits^1_0 dx \, x^N \,
g(x) \, .
\end{equation}

This consideration shows that a non-forward nature of hard exclusive reactions
does not complicate much their analysis in the high energy limit. Therefore
the method used in DIS \cite{Catani:1990xk,*Catani:1994sq} may be applied here.
The difference between DIS and our case is in the different form of $k_t$ dependent amplitudes
for corresponding partonic subprocesses.

Below I will concentrate on the dominant at high energy gluon contribution.
The results will be presented for the process (\ref{process}) ( assuming for simplicity
the asymptotic form of meson DA) and for the process of heavy quarkonium electroproduction
(where the formation of quarkonium is treated in NRQCD). The amplitude is presented as follows
\begin{equation}
\label{forcalc}
{\cal I}m A^g\sim {H^g(\xi,\xi)}
+ \int\limits^1_{2\xi}\frac{d x}{x}{H^g(x,\xi)}
\sum\limits_{n=1}C_n(L)\frac{\bar \alpha_s^n}{(n-1)!}\log^{n-1}\frac{x}{\xi} \, ,
\end{equation}
here $\bar \alpha_s=N_c \alpha_s/\pi$, we omitted normalization factors irrelevant for the subsequent discussion,
in the r.h.s $H^g(\xi,\xi)$ represents the Born contribution and the sum stands for the high
energy terms. $C_n(L)$ are the polynomials of variable $L=\log\frac{Q^2}{\mu_F^2}$ which we need to calculate.

Note that
the Born term belongs to the regular part (in terms of (\ref{DN})),
whereas the high energy terms behave as $(\bar \alpha_s/N)^n$ at $N\to 0$.
Therefore in the high energy terms one can replace gluon GPD in (\ref{forcalc})
by its forward limit, {$H^g(x,\xi)\to x g(x)$},  but
in the Born contribution
{$H^g(\xi,\xi)$} should be kept different from $x g(x)$.

Omitting all details of the derivation we just present the results. We work in $\overline{\rm{MS}}$ scheme.
We define (properly normalized)
$k_t$ dependent amplitude of the gluon subprocess
\begin{equation}
\label{liVM}
h_V(k_t^2)=\int\limits^1_0 dz \,
\frac{Q^2}{k_t^{\,\, 2}+z (1-z) Q^2}\phi_V (z)/\int\limits^1_0 dz \,
\frac{\phi_V (z)}{z(1-z)} \, ,
\end{equation}
then we  calculate its Mellin transform
\begin{equation}
\label{galiVM}
h_V(\gamma)=\gamma \int\limits^\infty_0\frac{d  k_t^{\,\, 2}}{k_t^{\,\, 2}}
\left(\frac{k_t^{\,\, 2}}{Q^2}\right)^\gamma h_V(k_t^2)=\frac{\Gamma^3[1+\gamma]\Gamma[1-\gamma]}{\Gamma[2+2\gamma]}\, .
\end{equation}
The high energy terms are defined from the expression
\begin{equation}
\label{char}
C_N^g\sim h_V(\gamma) R \left(\frac{Q^2}{\mu_F^2}\right)^\gamma \, .
\end{equation}
The gluon anomalous dimension is determined by the solution of equation $1=(\bar \alpha_s/N) \chi(\gamma)$,
where $\chi(\gamma)$ is the BFKL eigenfunction, function $R$ depends on $\bar \alpha_s/N$ and
is defined in \cite{Catani:1990xk,*Catani:1994sq}.
Expanding $C_N^g$ in the series of variable $y=\bar \alpha_s/N$ one can obtain
analytical expressions for the polynomials
$C_n(L)$.

Below we illustrate the values of these polynomials for the case $\mu_F=Q$
\begin{eqnarray}
V: &  1 - 2\, y + 4\, y^2 -2.39\, y^3 -4.09\, y^4 +\dots  & \nonumber \\
\rm{onium}: &  1 -1.39\, y +2.61\, y^2 +0.481\, y^3 -4.96\, y^4 +\dots & \nonumber \\
F_L: & 1 - 0.33\, y + 2.13\, y^2 +
    2.27\, y^3 + 0.434\, y^4 +\dots  & \nonumber
\end{eqnarray}
here the first two lines represent results for the exclusive light vector meson and quarkonium
production respectively, in the third line we show for comparison the results for longitudinal DIS structure function
\cite{Catani:1990xk,*Catani:1994sq}. We see that numerical values of $C_1(0)$
are negative in all case, but for the exclusive reactions its absolute values are about $4\div 6$ times larger
then in the case of $F_L$, explaining very large negative NLO corrections found for exclusive meson production.
On the other hand, the values of the second polynomial are positive and large,
$C_2(0)=4$ for the light vector meson production. This gives a hope that inclusion of these high energy terms
in the analysis may stabilize predictions for exclusive meson production.

To investigate this possibility we perform the following numerical study.
We calculate the amplitude of light vector meson production
with (\ref{forcalc}), where in the high energy terms we
use very simple input for the gluon density $
{H^g(x,\xi)}\sim x g(x)\sim x^{-0.2}$,
for the Born term we take ${H^g(\xi,\xi)}= \, 1.2 \, \xi\, g(\xi)$. Definitely, more realistic input
for gluon GPD should be used (especially for $H^g(\xi,\xi)$), but at the present
stage we just want to clarify the qualitative role of the high energy terms.
In Fig. 2 we present the energy dependence of the amplitude (in arbitrary units) calculated for
two values of photon virtuality $Q^2= 10$ and $20 \, \rm{GeV}^2$, for the running coupling we use
$\alpha_s(10)=0.25$ and $\alpha_s(20)=0.16$, and for the factorization scale $\mu_F^2=Q^2/2$.
The solid line on Fig. 2 represents the Born contribution, the dashed line -- the Born $+$ the first high energy term,
the dotted line -- the Born $+$ 2 first high energy terms, the dashed-dotted -- the Born $+$ 6 first high energy terms.
\begin{figure}[h!]
\begin{center}
\begin{tabular}{cc}
\mbox{\epsfig{figure=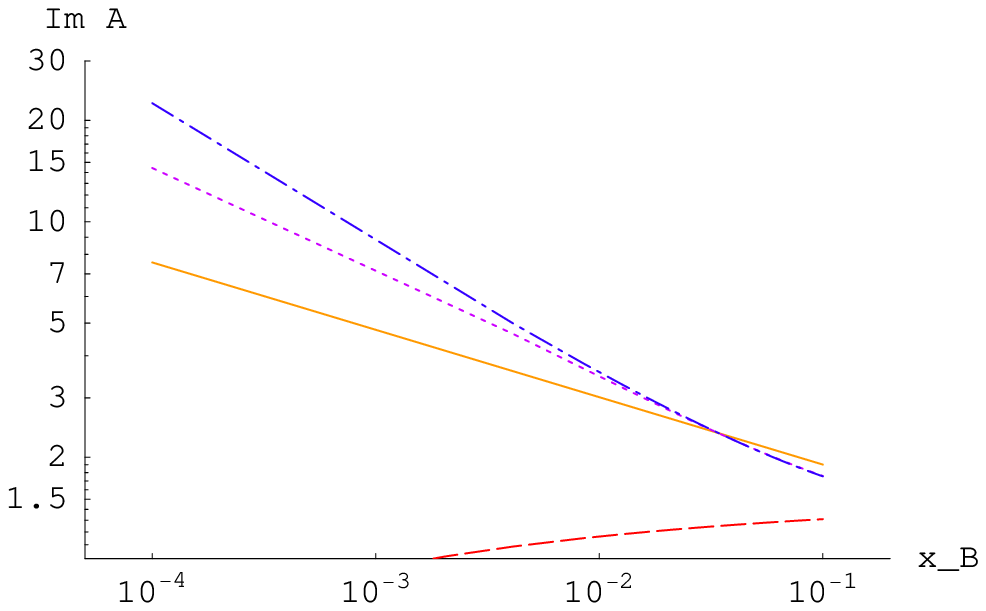,width=7.cm,height=5.cm}}&
\mbox{\epsfig{figure=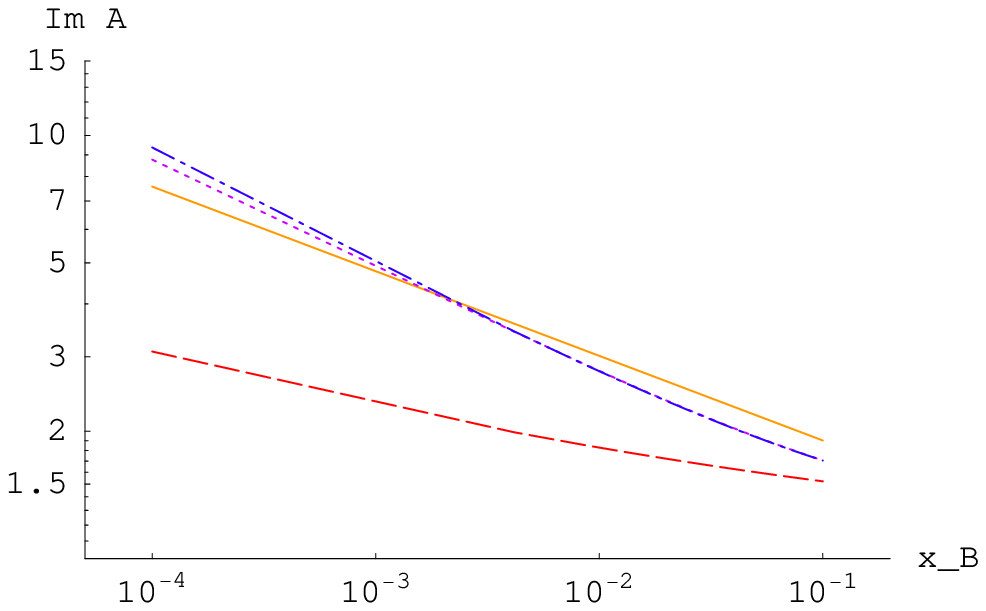,width=7.cm,height=5.cm}}\\
{\bf(a)}& {\bf(b)}
\end{tabular}
\end{center}
\caption{ The convergence of the high energy resummation:
{\bf(a)} $Q^2=10 \rm{GeV}^2$, {\bf(b)} $Q^2=20  \rm{GeV}^2$.  }
\end{figure}
We see that high energy resummation is convergent fast, the difference between the dashed and
the dashed-dotted lines is not big  for $Q^2= 10\, \rm{GeV}^2$ and is really
small for $20 \, \rm{GeV}^2$ cases. The other observation is that the inclusion of only first high energy term
(dashed line) seems to be a bad approximation. Even for $20 \, \rm{GeV}^2$ case, where
the Born and the resummed results are close to each other,  the dashed line is about factor of 3 below.

\begin{figure}[h!]
\begin{center}
\begin{tabular}{cc}
\mbox{\epsfig{figure=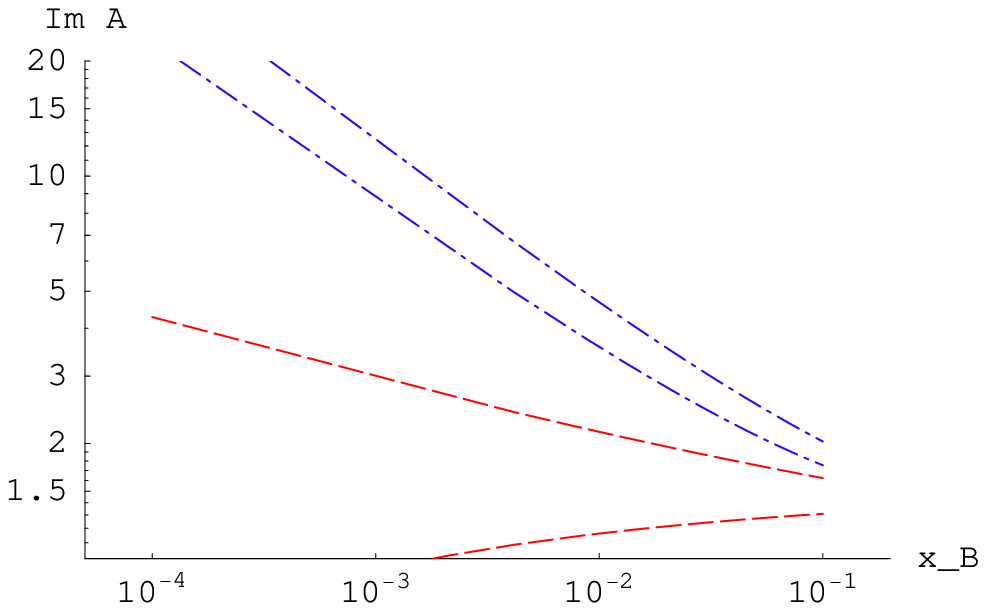,width=7.cm,height=5.cm}}&
\mbox{\epsfig{figure=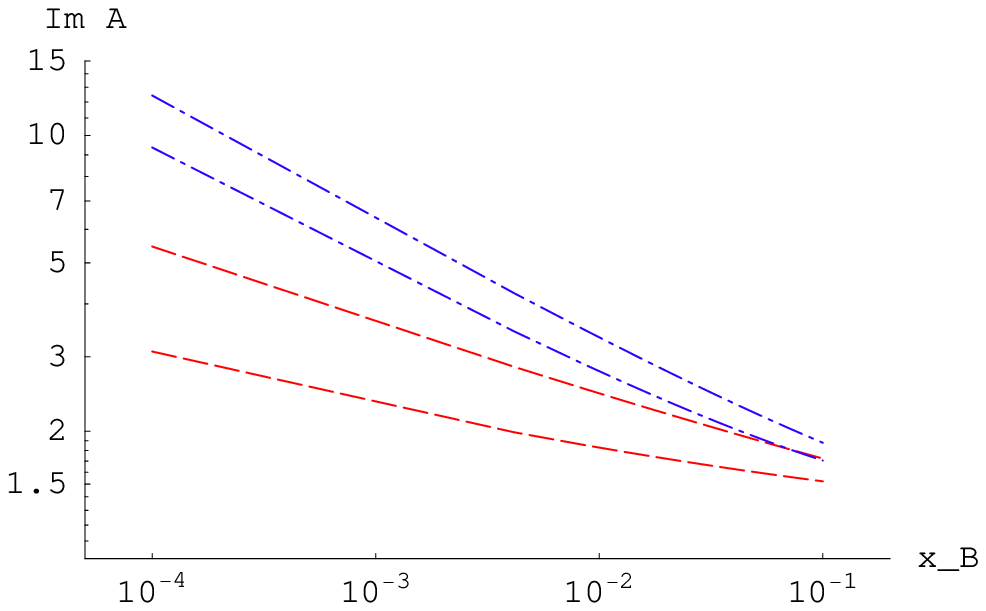,width=7.cm,height=5.cm}}\\
{\bf(a)}& {\bf(b)}
\end{tabular}
\end{center}
\caption{  {\bf(a)} $Q^2=10 \rm{GeV}^2$, {\bf(b)} $Q^2=20  \rm{GeV}^2$ }
\end{figure}

The dependence of the amplitude on the choice of factorization scale is shown in Fig. 3.
Again, the dashed lines correspond to the Born $+$ the first high energy term,
the dashed-dotted lines -- the Born $+$ 6 first high energy terms. The upper dashed and dashed-dotted lines
are for $\mu_F^2=Q^2/4$, the lower dashed and dashed-dotted lines
are for $\mu_F^2=Q^2/2$. We observe sizable reduction of the factorization scale dependence if
the high energy terms are resummed in comparison to the case when only the first of these terms
is taken into account.

\section{Summary}

Large NLO corrections are found for hard exclusive vector meson production.
At intermediate to larger values of $x_B$, typical for fixed-target experiments,
it seems that NLO corrections start to be under control for the large values of
$Q^2$, say above $10 \, \rm{GeV}^2$. However, the situation is much worse for the region of small $x_B$,
typical for the HERA collider experiments. Here NLO corrections are not under control even for
such large values of hard scales as $ 30 \, \rm{GeV}^2$, which prevents the interpretation of the precise
HERA data in terms of GPDs.
The problem is related to appearance of BFKL type logarithms in the hard-scattering amplitudes,
that calls for a resummation of these effects at higher orders.
Here we present the first results for such study.
The methods used
earlier for forward DIS process may be generalized to the case of nonforward hard exclusive reactions. We
obtained analitical results for the corresponding high energy terms in (\ref{forcalc}). The
first numerical calculation incorporating the high energy resummation is encouraging.

 \bigskip

\noindent {\it I am very grateful to the organizers and to Alexander von Humboldt Foundation for the
support of my participation in EDS07
conference. This work is also sponsored in part by grants
RFBR-06-02-16064 and NSh 5362.2006.2.}

\begin{footnotesize}
\bibliographystyle{blois07}
{\raggedright
\bibliography{blois07}
}
\end{footnotesize}
\end{document}